\documentclass[prl, aps,showpacs, notitlepage, groupedaddress,superscriptaddress, twocolumn, numerical]{revtex4-1}

\usepackage[utf8x]{inputenc}
\usepackage{color}
\usepackage{bbm}

\usepackage{nicefrac}
\usepackage{amsfonts,amsmath,amssymb,stmaryrd}
\usepackage{braket}

 \usepackage[autostyle=true]{csquotes}

\usepackage{tabularx}
\usepackage{multirow} 
\usepackage{hhline}
\usepackage{graphicx}
\usepackage{subfigure}  % use for side-by-side figures
\usepackage{bbm} 
\usepackage{hyperref}
\usepackage[pdftex]{epsfig}
\usepackage{mathrsfs}
\usepackage{verbatim}
\usepackage{centernot}
\usepackage{ulem}
\usepackage{array}
\usepackage{cancel}
\usepackage{ifthen}
\usepackage{bm}	% bold in math mode
\usepackage{siunitx}
\usepackage{todonotes}
\usepackage{setspace}
\usepackage{bm} % used for making vectors bold

\InputIfFileExists{gitinfo-latex.inc}{}{}

\usepackage{listings}
\usepackage{color}

\usepackage[acronym,nomain]{glossaries}

\begin{document}

\title{Single-atom quantum probes for ultracold gases using nonequilibrium spin dynamics}
\author{Quentin Bouton}
\affiliation{Department of Physics and Research Center OPTIMAS, Technische Universit\"at Kaiserslautern, Germany}

\author{Jens Nettersheim}
\affiliation{Department of Physics and Research Center OPTIMAS, Technische Universit\"at Kaiserslautern, Germany}
	
\author{Daniel Adam}
\affiliation{Department of Physics and Research Center OPTIMAS, Technische Universit\"at Kaiserslautern, Germany}
	
\author{Felix Schmidt}
\affiliation{Department of Physics and Research Center OPTIMAS, Technische Universit\"at Kaiserslautern, Germany}

\author{Daniel Mayer}
\affiliation{Department of Physics and Research Center OPTIMAS, Technische Universit\"at Kaiserslautern, Germany}
	
\author{Tobias Lausch}
\affiliation{Department of Physics and Research Center OPTIMAS, Technische Universit\"at Kaiserslautern, Germany}
	
\author{Eberhard Tiemann}
\affiliation{Institut f\"ur Quantenoptik, Leibniz Universit\"at Hannover, 30167 Hannover, Germany}	
		
\author{Artur Widera}
\email{email: widera@physik.uni-kl.de}
\affiliation{Department of Physics and Research Center OPTIMAS, Technische Universit\"at Kaiserslautern, Germany}
\affiliation{Graduate School Materials Science in Mainz, Gottlieb-Daimler-Strasse 47, 67663 Kaiserslautern, Germany}	

\date{\today}

\begin{abstract}
Quantum probes are atomic-sized devices mapping information of their environment to quantum mechanical states.  By improving measurements and at the same time minimizing perturbation of the environment, they form a central asset for quantum technologies. We realize spin-based quantum probes by immersing individual Cs atoms into an ultracold Rb bath. Controlling inelastic spin-exchange processes between probe and bath allows mapping motional and thermal information onto quantum-spin states.  We show that the steady-state spin-population is well suited for absolute thermometry, reducing temperature measurements to detection of quantum spin distributions. Moreover, we find that the information gain per inelastic collision can be maximized by accessing the nonequilibrium spin dynamic.  
The sensitivity of nonequilibrium quantum probing effectively beats the steady-state Cramér Rao limit of quantum probing by almost an order of magnitude, while reducing the perturbation of the bath to only three quanta of angular momentum. Our work paves the way for local probing of quantum systems at the Heisenberg limit, and moreover for optimizing measurement strategies via control of nonequilibrium dynamics.
\end{abstract}
\maketitle

\begin{figure}[t]
	\begin{center}
		\includegraphics[scale=0.70]{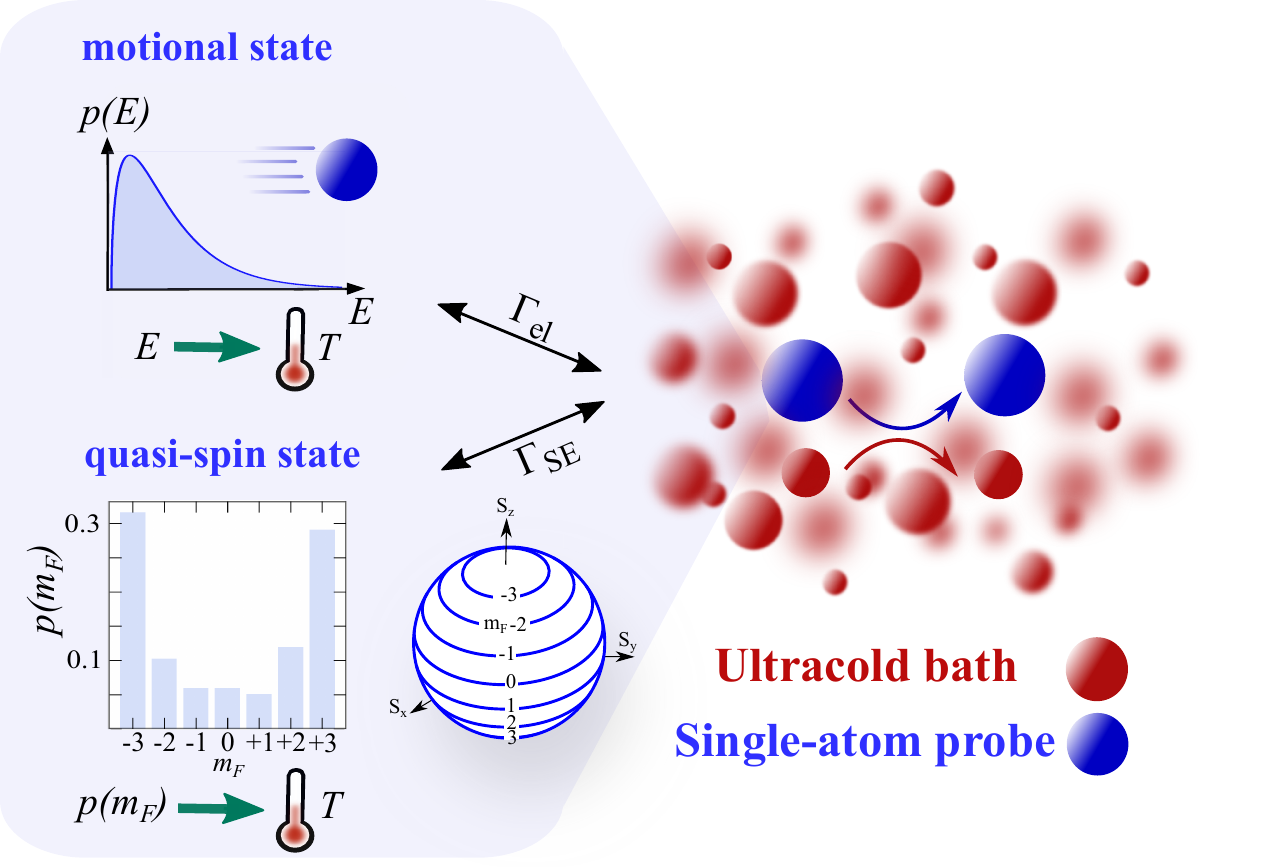}
	\end{center}
	\caption{\textbf{Single-atom quantum probing of an ultracold gas.} A single atom is coupled to the bath, interacting via two processes. First, elastic collisions at rate $\Gamma_\mathrm{el}$ thermalize the probe to the temperature of the bath, where the kinetic energy distribution allows for classical thermometry (top). Second, spin-exchange collisions at rate $\Gamma_\mathrm{SE}$ ensure motion-spin mapping and allow to store information on the bath energy in the quantum states of the probe, here the quasi-spin states, visualized by a macroscopic Bloch sphere (bottom).}
	\label{Figure_general_principle_thermometer}
\end{figure}

Miniaturizing measurement probes is a strong technological driving force and yields fascinating new insights into various fields including biology \cite{Kucsko2013}, solid-state physics \cite{Haupt2014} and metrology \cite{Kotler2011}.  A fundamental limit of miniaturization is the use of single atoms as individual probes, opening the door to employing quantum properties for advanced probing. A paradigm for quantum probing is a single atom with discrete energy quantum levels coupled to an atomic environment. 
Extracting relevant information stored in quantum levels of the probe can enhance the information obtained about a (quantum) environment under investigation.  
At the same time, the unavoidable perturbation of the environment caused by the measurement process can be reduced.  The potential of quantum probes has been at the focus of intense recent theoretical studies \cite{Degen2017,Johnson2011,Correa2015}, with a strong emphasis on quantum thermometry.  In classical thermometry, a thermometer thermalizes with the bath, and the mean kinetic energy of the probe is taken as a measure for the bath temperature presuming a Maxwell-Boltzmann distribution (Fig. \ref{Figure_general_principle_thermometer}). Thermometry of quantum systems is particularly important for ultracold gases, and various probes including magnons \cite{Olf2015}, confined Bose-Einstein condensate \cite{Lous2017}, Fermi sea \cite{Spiegelhalder2009} or single atoms \cite{Hohmann2016} have been reported. 
All these probes rely on the standard method of time-of-flight velocimetry \cite{Stamper1999} and thus are classical.  Exploiting the quantum properties of probes, however, has been shown to enhance precision and sensitivity, being ultimately limited by the Cramér Rao relation \cite{Helstrom1976}.  Numerous schemes have thus been proposed to extract temperature or work distributions via quantum probing \cite{Johnson2016,Degen2008,Johnson2011,Rivas2014,Retzker2008,Dorner2013}. The experimental demonstration of probing an atomic gas using the quantum properties of individual atoms, however, is so far elusive. Moreover, having access to the dynamics of the microscopic process of quantum probing opens the door to optimizing the information content obtained from the probe using nonequilibrium dynamic.\\

\begin{figure*}[t!]
	\begin{center}
		\includegraphics[scale=0.42]{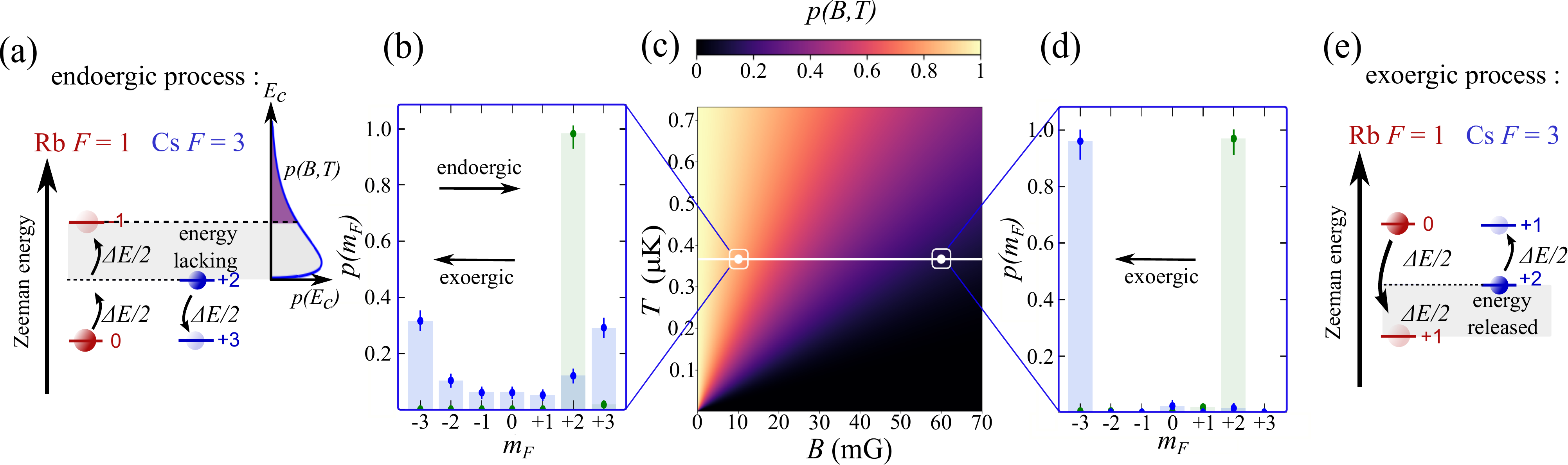}
	\end{center}
	\caption{\textbf{Mechanism of motion-spin mapping onto quantum states.} (a) Sketch of the endoergic process. Cs atoms go to a lower magnetic substate and release $\Delta E/2$ of energy, whereas Rb atoms go to a higher magnetic substate and require $\Delta E$ of energy. Therefore, this process can only occur if the missing Zeeman energy can be provided by kinetic energy $E_{c}$ during the collision, which is only possible for a fraction $p(B,T)$ of the probe atoms. (b) Experimental spin population of Cs atoms for a  magnetic field of $B=10\,$mG before (green) and after 350 ms interaction time in a Rb bath at $366^{+60}_{-40}$ nK (blue). 
%Since both processes are allowed ($p(B,T)=0.8$), 
We measure a nonzero population in $m_{F,\mathrm{Cs}}=+3$ (endoergic SE) and $m_{F,\mathrm{Cs}}=-3$ (exoergic SE). (c) Fraction of Cs atoms allowing to undergo an endoergic process as a function of the magnetic field $B$, assuming a Maxwell-Boltzmann distribution at temperature $T$. The indicated values correspond to measurements in (b) $p(B,T)=0.8$ and (d) $p(B,T)=0.1$; (d) Same as (b) but for $B=60\,$mG. Here, exoergic processes dominate, yielding a measured population (in blue) of Cs atoms in $m_{F,\mathrm{Cs}}=-3$. (e) Sketch of the exoergic process. Rb atoms are promoted to a lower magnetic state and release $\Delta E$ while Cs atoms are left in a higher magnetic state and only need $\Delta E/2$. As a consequence, this process is always allowed and releases $\Delta E/2$ of energy to the system.} 
	\label{figure_SE_iSE_process}
\end{figure*}

We realize a quantum probe using the discrete quasi-spin levels of a single Cesium (Cs) atom immersed in an ultracold gas of Rubidium (Rb) to store information about its temperature and also sensing the surrounding magnetic field. 
Moreover, we show that the sensitivity can be significantly enhanced by considering nonequilibrium spin dynamics of the quantum probe. The standard approach proposed for quantum probing is mapping of thermal information onto vibrational states of trapped particles \cite{Hangleiter2015} such as neutral atoms in optical tweezers \cite{Kaufman2012} or trapped ions \cite{Zipkes2010}.
Our approach of using quasi-spin states is particularly suited for ultracold temperatures and at the same time allows to independently control of trapping parameters. The relevant energy scales of our quantum probe are the thermal energy $k_B T$, with $k_B$ the Boltzmann constant, and the magnetic energy of the probe's Zeeman levels in a weak magnetic field $\Delta E/2 = g_{F} \mu_{B} B$, where $g_{F}$ is the Landé factor and $\mu_{B}$ the Bohr magneton.  For a magnetic field of $B=10$ mG, the energy splitting corresponds to $g_{F} \mu_{B} B / k_{B} \sim 170 $ nK. For comparison, this energy corresponds to a trap level spacing of $3.5\,$kHz, which is well below  values of vibrational level spacing for tight traps.\\

Individual laser-cooled Cs atoms are initially prepared in the Zeeman state $\ket{F_{\mathrm{Cs}}=3,m_{F,\mathrm{Cs}}=2}$, where $F_{\mathrm{Cs}}$ is the total angular momentum and $m_{F,\mathrm{Cs}}$ its projection on the quantization axis. 
The Rb bath is produced in $\ket{F_{\mathrm{Rb}}=1,m_{F,\mathrm{Rb}}=0}$ with temperatures ranging from $T=0.2$ to 1$\,\mathrm{\mu}$K \cite{Mayer2018,Supplemental_Material}. Interaction between probe atom and bath is initiated by transporting the Cs atom into the Rb cloud, and comprises two processes (Fig.~\ref{Figure_general_principle_thermometer}). First, frequent elastic collisions at rate $\Gamma_\mathrm{el}$ between probe atom and bath ensure thermalization of the probe's motional degree of freedom with the bath, while leaving the internal states unaffected. Second, motion-spin mapping is achieved via inelastic spin-exchange (SE) collisions at rate $\Gamma_\mathrm{SE} \approx \Gamma_\mathrm{el}/10$. SE collisions exchange individual quanta of angular momentum between probe and bath, where the Zeeman energy shifts for Rb and Cs differ by a factor of two due to $g_{F,\mathrm{Rb}} = 2 g_{F,\mathrm{Cs}}$. For exoergic (endoergic) SE an energy of $\Delta E/2$ is released (lacking) between initial and final states of the Cs-Rb collision partners while changing the atomic quasi-spin accordingly (see Fig.~\ref{figure_SE_iSE_process}). Exoergic processes are thus always allowed and tend to drive the probe's spin population toward $m_{F,\mathrm{Cs}} = -3$. By contrast, endoergic processes can only occur, if the missing energy difference of Zeeman states can be provided by the kinetic, and thus thermal, energy in the collisional process. This discrimination of SE by thermal energy is the microscopic mechanism of motion-spin mapping and effectively cools the collision partners similar to Pomeranchuck cooling \cite{Pobell2007}. In both SE processes, frequent elastic collisions quickly rethermalize the probe well before the next SE collision. The precise values of the SE rates depend on the atomic states as well as the full collisional energy and can be precisely modeled \cite{Supplemental_Material,Schmidt2019}. Important insight on the quantum probing, however, can be obtained from a purely energetic argument. The fraction of atoms that are energetically allowed to undergo an endoergic collision is given by \cite{Supplemental_Material}
\begin{displaymath}
p(B,T) = \int_{\Delta E/2}^{\infty} p(E_{c}) dE_{c},
\end{displaymath}

\noindent assuming a Maxwell-Boltzmann distribution $p(E_c)$ of collision energies $E_c$ (Fig.~\ref{figure_SE_iSE_process}). Therefore, modifying the relative contributions of Zeeman and thermal energies allows to microscopically tune the probability for an endoergic collision. Hence, the Cs spin distribution and its dynamics reflect precisely the competition between magnetic and thermal energies via the probability for endoergic collisions. In fact, any additional mechanism shifting the total energy of an atomic collision can also be sensed by our atomic quantum probe.\\

\begin{figure}[h!]
	\begin{center}
		\includegraphics[scale=0.45]{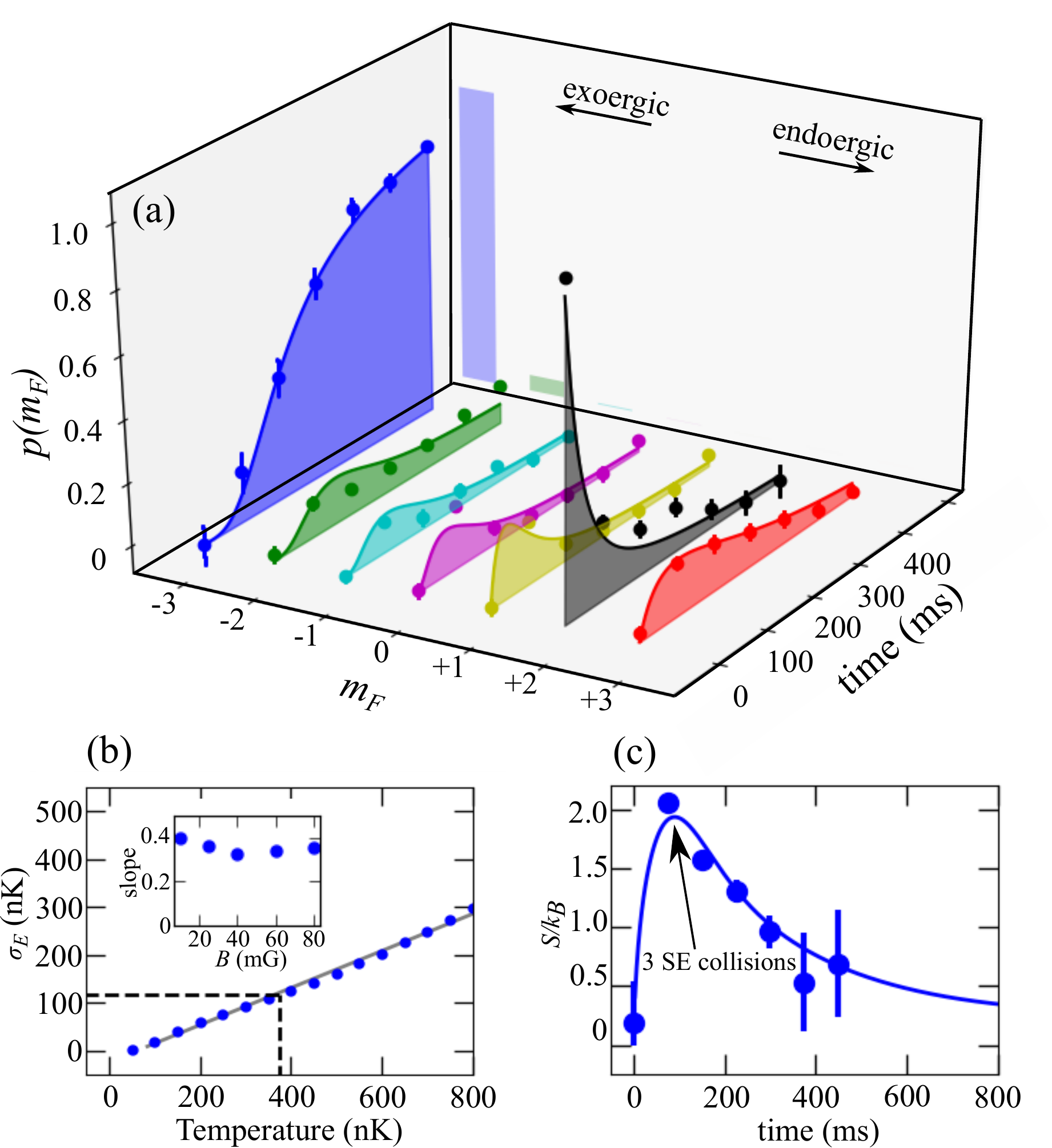}
	\end{center}
	\caption{\textbf{Information gain from the spin distribution of quantum probes.} (a) Quasi-spin states dynamics of Cs atoms immersed in a Rb bath at T = $366^{+60}_{-40}$ nK at a fixed magnetic field $B=25$ mG (expected endoergic fraction $p(B,T)=0.54$). Experimental time trace (dots) and theoretical predictions of our rate model \cite{Supplemental_Material} (solid lines) are shown. Each experimental point is an average over approximately 200 measurements, and the errorbars indicate the  statistical uncertainties in the atom number determination. The histogram projected in the back plane shows the steady-state spin distribution. (b) Theoretically calculated probe-energy fluctuations $\sigma_{E}^{2}$ of the steady-state for $B=25$ mG, calculated from modelled quantum-state distributions as $\langle E^{2} \rangle = \sum_{m_F} E^{2}_{m_F} p(m_F)$, with $E_{m_F}$ and $p(m_F)$ being the energy and population probability of quantum state $m_F$. The $p(m_F)$ are inferred finding the steady-state of our rate model \cite{Supplemental_Material} that in this case depend only on the scattering cross sections. Dashed lines represent $T=366$ nK. Inset: Slope of the linear trend between $\sigma_{E}$ and $T$ for different values of the magnetic field. (c) Time evolution of the probe entropy, calculated from the spin populations of (a). The time of maximal entropy corresponds to an average of three SE collisions (0.5 endoergic and 2.5 exoergic).}
	\label{figure_time_evolution_data}
\end{figure}

The ensuing time evolution of our quantum probe's spin population is shown in Fig. \ref{figure_time_evolution_data} (a) together with the projected steady-state spin distribution. We observe a redistribution of the probe's spin population over time toward the steady-state, due to the competition of the rates between the exoergic and endoergic processes. To reach this state, the probe has to undergo a dozen of SE collisions. Each SE collision also modifies the spin state of one Rb atom. 
The strong imbalance between the probe and the bath, and the relatively short interaction time, imply that the assumption of an ideal Markov bath applies here, $i.e.$ in every SE collision the probe interacts with a Rb atom in the initial quantum state $m_{F,\mathrm{Rb}}=0$. We model the time evolution of the probe's spin population with a full rate model. All SE processes are integrated, based on high-precision data at ultralow atomic collision energies obtained in previous work \cite{Schmidt2019}. In short, the SE collision rates $\Gamma = \braket{n} \sigma \bar{v}$ of the model are directly inferred from atomic cross sections $\sigma$, where $\bar{v}$ is the relative velocity between Rb and Cs, and $\braket{n}$ their density overlap, both calculated assuming thermalized atoms \cite{Supplemental_Material}. The different values of the scattering cross sections $\sigma$ and their dependence on temperature $T$ and magnetic field $B$ are based on a precise model of the Rb-Cs molecular potential \cite{Takekoshi2012,Supplemental_Material,Schmidt2019}. Our rate model fully captures the spin dynamics and yields excellent agreement for the time evolution of the probe's spin population for all parameters.    \\

Absolute bath thermometry can be performed using the probe's steady-state quasi-spin distribution. For the limiting case approaching $T=0$, endoergic processes are absent, and the steady-state is a polarized state of the probe in $m_{F,\mathrm{Cs}}=-3$. For increasing temperature, endoergic processes  emerge, leading to a spreading of the quantum probe's steady-state spin population. We thus investigate the fluctuations of the energy associated with the probe's steady-state spin population  $\sigma^{2}_{E} = \braket{E^{2}}- \braket{E}^{2}$ for different bath temperatures, shown in Fig. \ref{figure_time_evolution_data} (b). We find a linear increase of the spin distribution's width with bath temperature, where the proportionality constant is independent of the specific magnetic field value, but also of the initial state of the probe, Rb densities and number of spin collisions since we consider here the steady-state. Hence, our quantum probe is well suited for absolute thermometry, allowing to extract temperature information from spin-population measurements at known magnetic field values. Importantly, the steady-state observed is not the equilibrium state of the total system. After $\sim 10^3$ SE collisions, the Markov approximation will break down, and the spin-states of the Rb bath will significantly change toward the global equilibrium state. This regime is experimentally not accessible and thus neglected.\\

While steady-state thermometry yields information which is independent of the details of the interaction, experimentally it features several drawbacks. First, atom loss can prevent long interaction times, especially for large bath densities.  Second, albeit the number of SE collisions is small compared to the number of atoms in the gas, identifying the least-perturbative measurement protocol for quantum probing is of fundamental interest. We therefore investigate the information obtained during the nonequilibrium time evolution of the probe's spin distribution. To quantify the information gain per SE collision, we plot the time dependence of the Shannon entropy \cite{Shannon48} of the quantum probe's spin distribution $S = -k_{B} \sum_{m_{F}} p(m_{F}) \mathrm{log} (p(m_{F}))$ in Fig.~\ref{figure_time_evolution_data}(c). We find a maximum of the entropy for only three SE collisions, indicating that, for the initial conditions used, the nonequilibrium spin distribution can provide much more information than the steady-state while minimizing the bath perturbation. \\

\begin{figure}[t!]
	\begin{center}
		\includegraphics[scale=0.425]{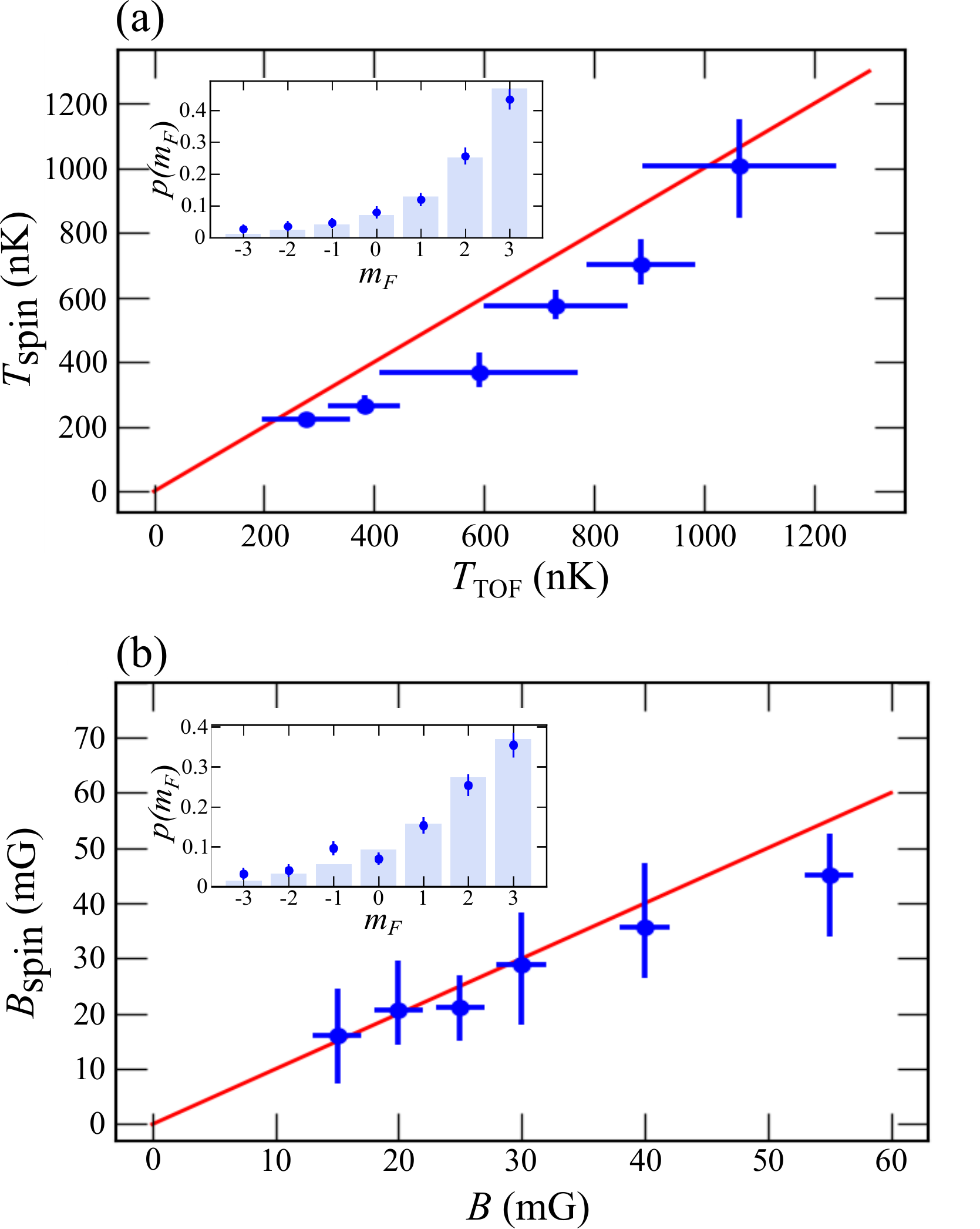}
	\end{center}
	\caption{\textbf{Nonequilibrium quantum probing.} (a) Comparison of the temperatures extracted from the spin population of Cs atoms $T_{\mathrm{spin}}$ to time-of-flight temperatures of the Rb cloud $T_{\mathrm{TOF}}$ for a fixed magnetic field $B=10$ mG. The errorbars of $T_{\mathrm{spin}}$ originate from the statistical errors on the $\chi^{2}$-analysis and the errorbars of $T_{\mathrm{TOF}}$ reflects the shot-to-shot fluctuations in the experiment (around $10 \% - 15 \% $). The red line serves as a guide to the eye $T_{\mathrm{spin}}$ = $T_{\mathrm{TOF}}$. Inset: Example of spin population. The dots represent the data and the histogram the theory with the best fitting temperature, here $T_{\mathrm{spin}} = 1008^{+140}_{-160}$ nK. (b) same as (a) but with the magnetic field $B$ for a fixed temperature $T=1 \mu $K. Inset : Spin population with $B_{\mathrm{spin}}=20.7^{8.9}_{-6.2}$ mG.}
	\label{figure_T_ToF_vs_Tspin}
\end{figure}

We quantify the performance of the nonequilibrium probing by first considering the information obtained from finite-time data, taken at an interaction time of 350 ms. We perform thermometry or magnetometry by varying bath temperature or magnetic field value, leaving the respective other value fixed. The nonequilibrium values for temperature ($T_\mathrm{spin}$) or magnetic field ($B_\mathrm{spin}$) are determined by comparing the measured quasi-spin populations with our numerical model using a $\chi^{2}$-analysis \cite{Supplemental_Material,Bevington2013}, where only $T_\mathrm{spin}$ or $B_\mathrm{spin}$ is a free parameter. We compare the quantum probe's values with independently measured values of time-of-flight velocimetry for temperature, and microwave spectroscopy of Rb hyperfine transitions for magnetic field. We find in general good agreement, despite the fact that, in temperature measurements for instance, motional information of the Rb bath is compared to spin-based information of the Cs quantum probe. \\

Second, we investigate the sensitivity of the nonequilibrium probing, making use of the Quantum Fisher information $F$ as an indicator of the thermal and magnetic sensitivities. 
Fisher information is a key concept in parameter estimation theory \cite{Braunstein1994} and has been used to quantify many observables, ranging from temperature to entanglement \cite{Correa2015,Wasilewski2010,Boss2017,Strobel12017}, and recently for cold atom magnetometry \cite{Evrard19}. Neglecting coherence in the system, we describe each state by a diagonal density matrix $\hat{\rho}(B,T)$=$\sum_{m_{F}} P_{m_{F}}(B,T)  \ket{m_{F}} \bra{m_{F}}$, where $P_{m_{F}}(B,T)$ are the spin populations of the probe at $T$ and $B$. We denote the parameter of interest as $\theta$ ($\theta=B$ or $T$). We quantify the distance between two quantum states at $\theta$ and $\theta + \delta \theta$ using the Bures distance as \cite{Supplemental_Material} \\
\begin{equation}
\label{Bures_distance_definition}
d^{2}_{\mathrm{Bures}}(\delta \theta) = 2-2 \sum_{m_{F}} \left[ P_{m_{F}}(\theta)P_{m_{F}}(\theta+\delta \theta ) \right]^{1/2},
\end{equation}

\begin{figure}[t]
	\begin{center}	
		\includegraphics[scale=0.5]{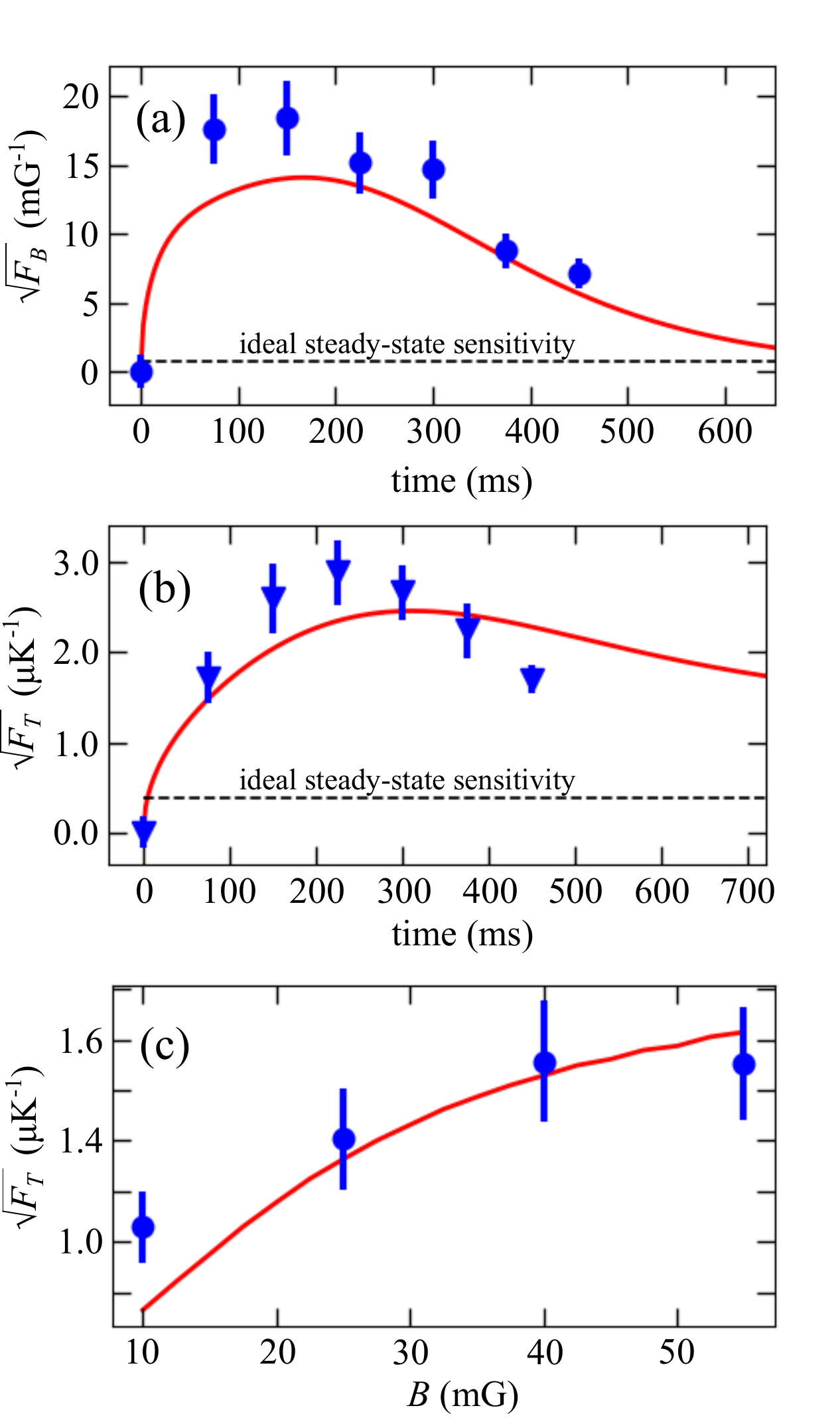}
	\end{center}
	\caption{\textbf{Nonequilibrium boost of quantum-probe sensitivity.} Time dependence of the magnetic sensitivity $\sqrt{F_{B}}$ (for $T=590\,$nK, centred at $B=40\,$mG ) (a) and thermal sensitivity $\sqrt{F_{T}}$ (for $B=40\,$mG, centred at $T=640\,$nK) (b) of the quantum probe. Red lines correspond to theoretical calculations, blue dots/triangles are experimental data; dots are sensitivities extracted comparing only experimental populations whereas triangle indicate sensitivities extracted comparing measured populations to theoretical ones. Dashed lines represent the respective sensitivity value expected for the steady state. (c) Thermal sensitivity (centred at $T$=640 nK) as a function of the $B$ field for an interaction time that fixes the number of SE collisions to approximately 4.}
	\label{figure_sensitivity}
\end{figure}

\noindent which coincides with the Hellinger distance \cite{Luo2004} for commuting density operators. A Taylor expansion to first order of the Bures distance defines the usual connection between Bures distance and $F_{\theta}$ \cite{Supplemental_Material}
\begin{equation}
\label{Bures_distance_definition_T}
 d_{\mathrm{Bures}}(\delta \theta) = \sqrt{F_{\theta}}  \delta \theta \ + \ \mathcal{O} (\delta \theta)^{2}.
\end{equation}

\noindent Hence, high sensitivities, indicated by a large value of $F_{\theta}$, also imply a high statistical speed $\partial d_{\mathrm{Bures}} / \partial \delta_{\theta} = \sqrt{F_{\theta}}$ to change the Bures distance according to the parameter change. Thus we will refer to $\sqrt{F_{\theta}}$ as sensitivity. 
First, we investigate the time evolution of the thermal ($\theta=T$) and magnetic ($\theta=B$) sensitivities of our quantum probe (Fig  \ref{figure_sensitivity} (a) and (b)). 
We observe that the sensitivity reaches a maximum  in both cases, which outperforms the steady-state sensitivity by a factor of 6.55 (17.5) for thermometry (magnetometry). 
This implies that nonequilibrium probing also outperforms the Cramér-Rao bound \cite{Helstrom1976} of steady-state probing. In both cases, this time is close to the time where the entropy of the quantum probe's spin distribution is also maximum, \textit{i.e.~}where the amount of information gain is largest. Second, in Fig.~\ref{figure_sensitivity} (c), we study the thermal sensitivity of the probe at fixed time, adjusted to a constant number of 4.2(3) exoergic spin collisions. We observe that the sensitivity per collision increases with magnetic field $B$. This observation is explained by the decrease of the number of endoergic processes from $2.3(2)$ for $B$=10 mG to $0.5 (1)$ for $B$=55 mG ($T$ being fixed). The information about bath temperature is contained in the endoergic process. Hence, if the probability to undergo such a process is low, the amount of information carried by a single event increases, giving rise to a large information gain per endoergic collision. \\

The realization of individual atomic quantum probes yielding access to information obtained by nonequilibrium dynamics opens a new way to optimize quantum probing strategies, where our work has already demonstrated a boost of sensitivity of roughly an order of magnitude. Moreover, reducing the bath size will allow following the transition from a Markov to a non-Markov bath, shedding new light on the microscopic quantum dynamics for system-bath entanglement \cite{Breuer2009}. Finally, our experimental system also paves the way to local probing of quantum gases or employing collective interaction effects \cite{Mehboudi2019}. \\

We thank Eric Lutz for helpful discussions. This work was funded in the early stage by the European Union via ERC Starting grant "QuantumProbe" and in the final stage by Deutsche Forschungsgemeinschaft via Sonderforschungsbereich (SFB) SFB/TRR185.

%%%%%%%%%%%%%%%%%%%%%%%%%%%%%%%%%%%%%%%%%%%%%%%%%%%%%%%%%%%%%%%%%%%%%%%%%%%%%
%\bibliographystyle{apsrev4-1}

\section{}
\newpage
\setcounter{equation}{0}
\section{SUPPLEMENTAL MATERIAL}
\subsection{Experimental procedure}

The Rb cloud is prepared by loading a laser-cooled cloud into a crossed dipole trap at $\lambda =$ 1064 nm. Changing the final dipole trap depth at the end of the evaporation, we can create Rb clouds with typically $N = 5-9 \times 10^{3}$ atoms numbers and temperatures between 0.2 and 1 $\mu$K. The dipole trap is then adiabatically compressed to a fixed final trap depth, yielding trap frequencies in radial and axial directions of $\omega_{r}=2 \pi \times 330$ Hz and $\omega_{z}=2 \pi \times 50$ Hz respectively, and atomic densities on the order of $10^{12}-10^{13}$ $\mathrm {cm^{-3}}$. The Rb cloud is then transferred into the insensitive magnetic field state $m_{F,\mathrm{Rb}}=0$ by microwave sweeps. Subsequently, few Cs atoms are captured in a high-gradient magneto-optical trap and loaded into an independent crossed dipole trap, located at 160 $\mu$m from the Rb cloud. Cs atoms are further cooled down with a degenerate Raman side-band cooling scheme \cite{Kerman2000}, pumping the Cs atoms in their absolute ground state $m_{F,\mathrm{Cs}}=3$. Thereafter, the Cs atoms are transferred into the desired internal state $m_{F,\mathrm{Cs}}=2$ by microwave-driven Landau-Zener transitions, near-resonant to the hyperfine transition ($h$ $\times$ 9.1 GHz). The use of few Cs atoms (6 in average) is a compromise between neglecting Cs-Cs interactions \cite{Chin2004} and minimizing the influence on the bath on the one hand, and obtaining sufficient statistics on the other hand. The limit of single probes, however, is routinely possible. Finally Cs atoms are guided by the dipole trap potential into the ultracold cloud, before the interaction starts. Due to favourable ratio of mass and dipole force, Cs atoms experience almost the same trapping frequencies as Rb atoms. The magnetic field amplitude $B$ during the Cs-Rb interaction is calibrated with Rb atoms, using the $h$ $\times$ 6.8 GHz microwave transition that is resonant with the ground-level hyperfine splitting. This allows us to control $B$ with an accuracy of $\pm$ 2 mG. Moreover, we take care to ramp up the magnetic field adiabatically in order to avoid mixing Zeeman states. After an interaction time, the Rb cloud is removed from the trap by a resonant laser pulse. The populations of Cs atoms in the different $m_{F,\mathrm{Cs}}$ states are then infered by a combination of state-sensitive microwave transitions at $h$ $\times$ 9.1 GHz and a hyperfine sensitive push-out laser pulse. \cite{Schmidt2018}. 

\begin{figure}[t]
	\begin{center}
		\includegraphics[scale=0.45]{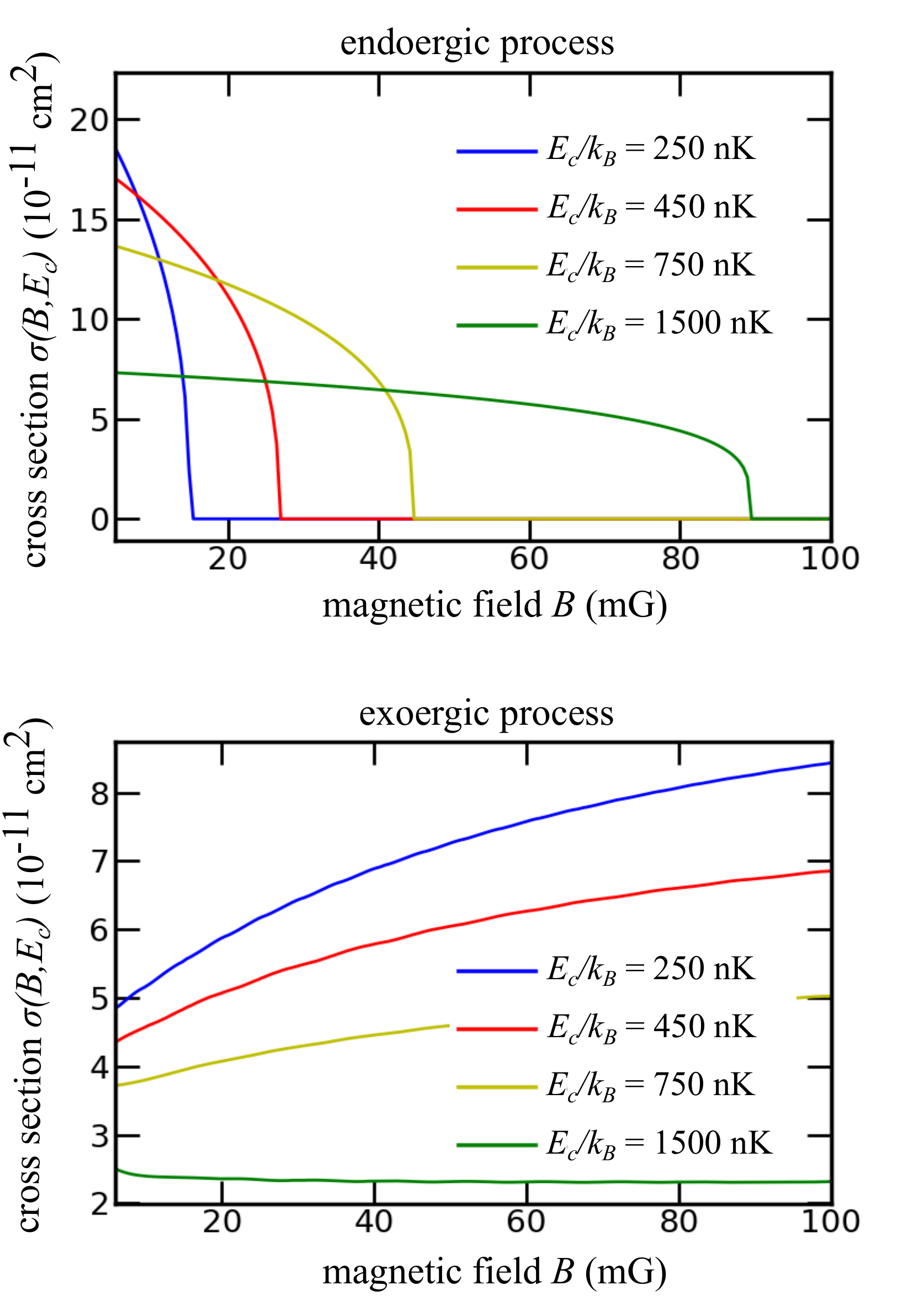}
	\end{center}
	\caption{Scattering cross sections for endoergic SE (top) and exoergic SE (bottom) for Cs in the state $m_{F,\mathrm{Cs}}=2$ and Rb in the state $m_{F,\mathrm{Rb}}=0$. The cross sections are plotted for 4 different fixed collisions energies $E_{c}$ (250 nK, 450 nK, 750 nK, 1500 nK). In the endoergic SE, the energy condition is underlined: if $E_{c} \leq \mu_{B} B/4$, the collision is forbidden and therefore the cross section drops to 0.}
	\label{figure_cross_section_Ec_supplementary}
\end{figure}

\begin{figure}[t]
	\begin{center}		
		\includegraphics[scale=0.45]{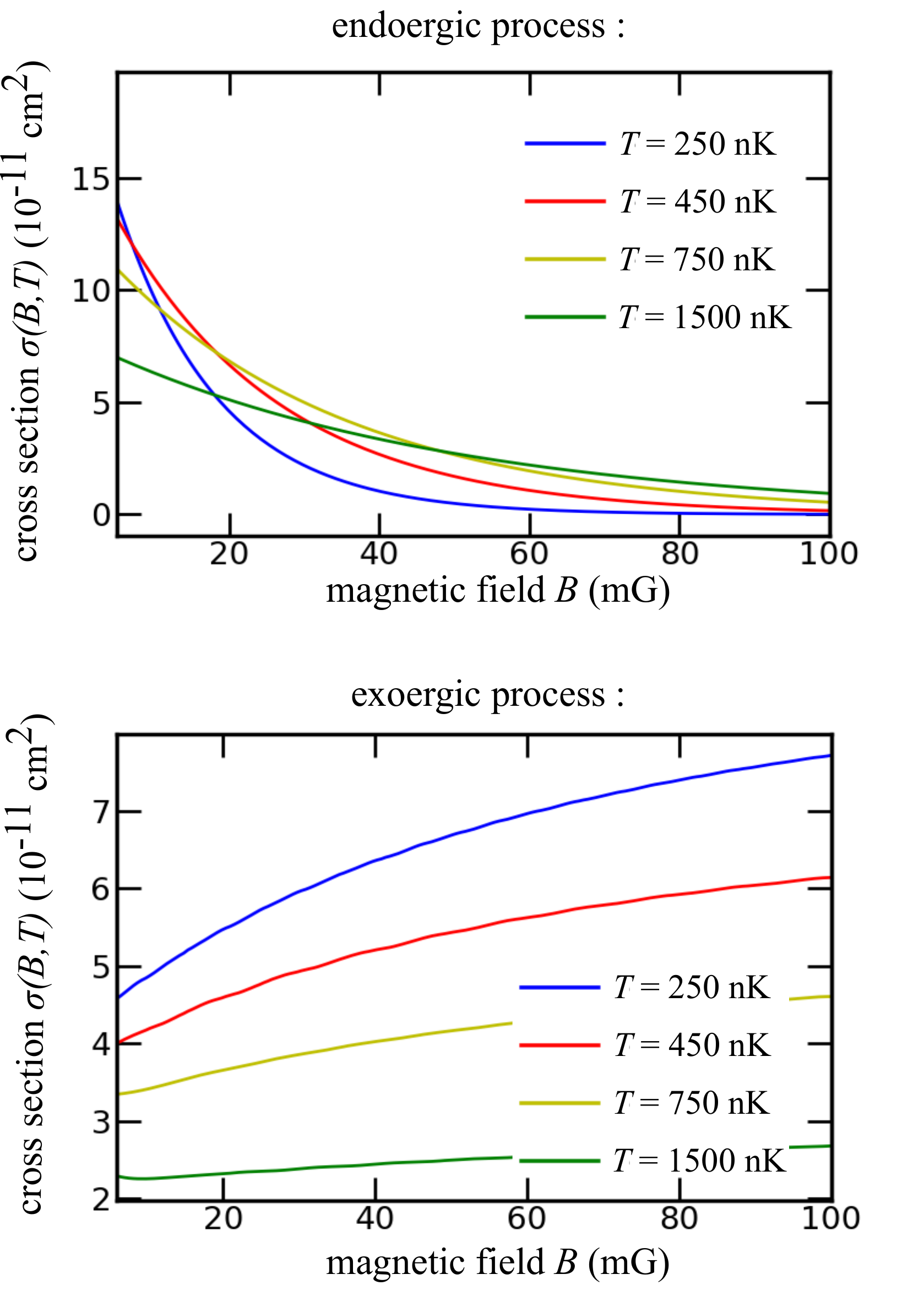}
	\end{center}
	\caption{Scattering cross sections for endoergic SE (top) and exoergic SE (bottom) for Cs in the state $m_{F,\mathrm{Cs}}=2$ and Rb in $m_{F,\mathrm{Rb}}=0$. The cross sections shown here include the effect of finite temperature and are plotted for 4 different temperatures $T$ (250 nK, 450 nK, 750 nK, 1500 nK).}
	\label{figure_cross_section_MB_supplementary}
\end{figure}

\subsection{Fraction of Cs atoms allowing to undergo an endoergic process}

During an endoergic spin exchange (SE) collision, the Cs atom delivers $\Delta E/2$ of energy, where $\Delta E/2$ is the Zeeman energy splitting of the Cs atom (See Fig. \ref{figure_SE_iSE_process} (a)). Operating at low magnetic field $B$, the splitting writes $\Delta E/2 = g_{F} \mu_{B} B$, where $g_{F}$ is the Landé factor ($g_{F}$=1/4), $\mu_{B}$ the Bohr magneton and $B$ the magnetic field. However, for the endoergic collision to occur, the Rb atom requires $\Delta E$ of energy. As a consequence, $\Delta E/2$ is lacking, which must be provided by the kinetic energy of the collision, which is given by $E_{c}= \mu v_{rel}^{2}/2$, where $\mu$ is the reduced mass of Rb and Cs and $v_{rel}$ their relative velocity. Assuming that Cs and Rb atoms are thermalized at temperature $T$, the collision energy $E_{c}$ follows a Maxwell-Boltzmann distribution \cite{Cannoni2014} 
\begin{equation}
\label{1}
p(E_{c}) = \left( \frac{1}{\pi k_{B}T} \right) ^{3/2} 2 \pi \sqrt{E_{c}} \exp \left(- \frac{E_{c}}{ k_{B} T} \right) 
\end{equation}

\noindent where $k_{B}$ is the Boltzmann constant. The fraction of Cs atoms allowing to undergo a SE is thus given by $p(B,T) = \int_{\Delta E/2}^{\infty} p(E_{c}) dE_{c}$ and writes
\begin{equation}
\label{2}
p(B,T) =1+ \sqrt{\frac{\mu_{B}B}{\pi k_{B}T}} \exp \left( \frac{\mu_{B}B}{4 k_{B} T} \right) - \mathrm{erf} \left( \sqrt{\frac{\mu_{B}B}{4 k_{B}T}} \right)  
\end{equation}

\noindent where erf is the error function. This is the expression used to plot the Fig. \ref{figure_SE_iSE_process} (c) in the main text.

\subsection{SE scattering cross sections}

The interaction between the Rb and Cs atoms is modelled by a molecular potential arising from the inter-particle singlet and triplet potentials. It allows for elastic and spin-exchange collisions \cite{Schmidt2019}. Elastic collisions preserve the internal states of both collision partners after the collisions, leading to thermalisation. SE processes lead to a spin transfer  while maintaining the total magnetization $M=m_{F,\mathrm{Rb}}+m_{F,\mathrm{Cs}}$. Thus only exoergic and endoergic processes with a spin transfer $\Delta m_{F,\mathrm{Cs}}= \pm 1$ for Cs and $\Delta m_{F,\mathrm{Rb}}= \mp 1$ for Rb are possible, where $\Delta m_{F,\mathrm{Cs}} < \Delta m_{F,\mathrm{Rb}}$ gives an endoergic process and $\Delta m_{F,\mathrm{Cs}} > \Delta m_{F,\mathrm{Rb}}$ an endoergic process. Scattering cross sections for respective SE processes are calculated in a coupled-channel scattering model. The calculations are based on a Cs-Rb interaction potential model, obtained from more than $30 \times 10^{3}$ spectroscopy lines and Feshbach resonances \cite{Takekoshi2012}. Each individual calculation uses a fixed magnetic field $B$ and a fixed collision energy $E_{c}$. They have been performed for all possible asymptotic channels $\ket{m_{F,\mathrm{Cs}},m_{F,\mathrm{Rb}}}$. In figure \ref{figure_cross_section_Ec_supplementary}, we show the calculated cross section $\sigma(B,E_{c})$ of the endoergic and exoergic process for Cs atoms initially in $m_{F,\mathrm{Cs}}=2$ and Rb in $m_{F,\mathrm{Rb}}=0$. \\

Additionally, we take into account the effect of the finite temperature in the cross section by calculating $\sigma(B,T) = \int p(E_{c}) \sigma(B,E_{c}) dE_{c}$, where $p(E_{c})$ is defined in Eq.~\eqref{1}. In figure \ref{figure_cross_section_MB_supplementary}, we plot the finite temperature cross sections of both SE processes for Cs atoms initially in $m_{F,\mathrm{Cs}}=2$ and Rb in $m_{F,\mathrm{Rb}}=0$. 

\begin{figure}[t]
	\begin{center}
		\includegraphics[scale=0.75]{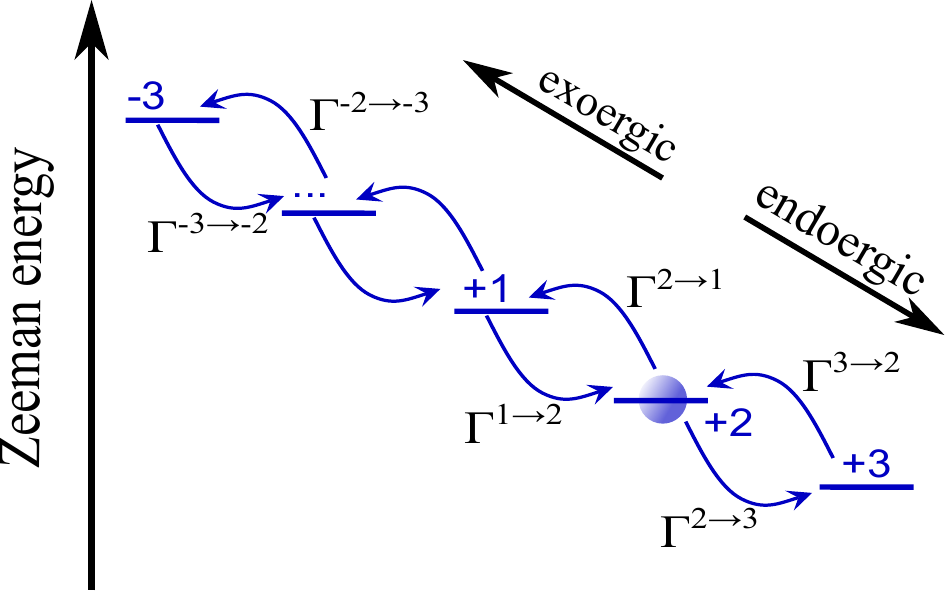}		
	\end{center}
	\caption{Sketch of the rate model used to model the spin dynamic. For each spin state $m_{F,\mathrm{Cs}}$, endoergic processes lead to a decay to the $m_{F+1,\mathrm{Cs}}$ state and a gain from the $m_{F-1,\mathrm{Cs}}$ state. Besides Exoergic SE processes lead to a decay to the $m_{F-1,\mathrm{Cs}}$ state and a gain from the $m_{F+1,\mathrm{Cs}}$ state.}
	\label{figure_rate_supplementary}
\end{figure}

\subsection{Spin-exchange model}

\begin{figure}[t]
	\begin{center}
		\includegraphics[scale=0.5]{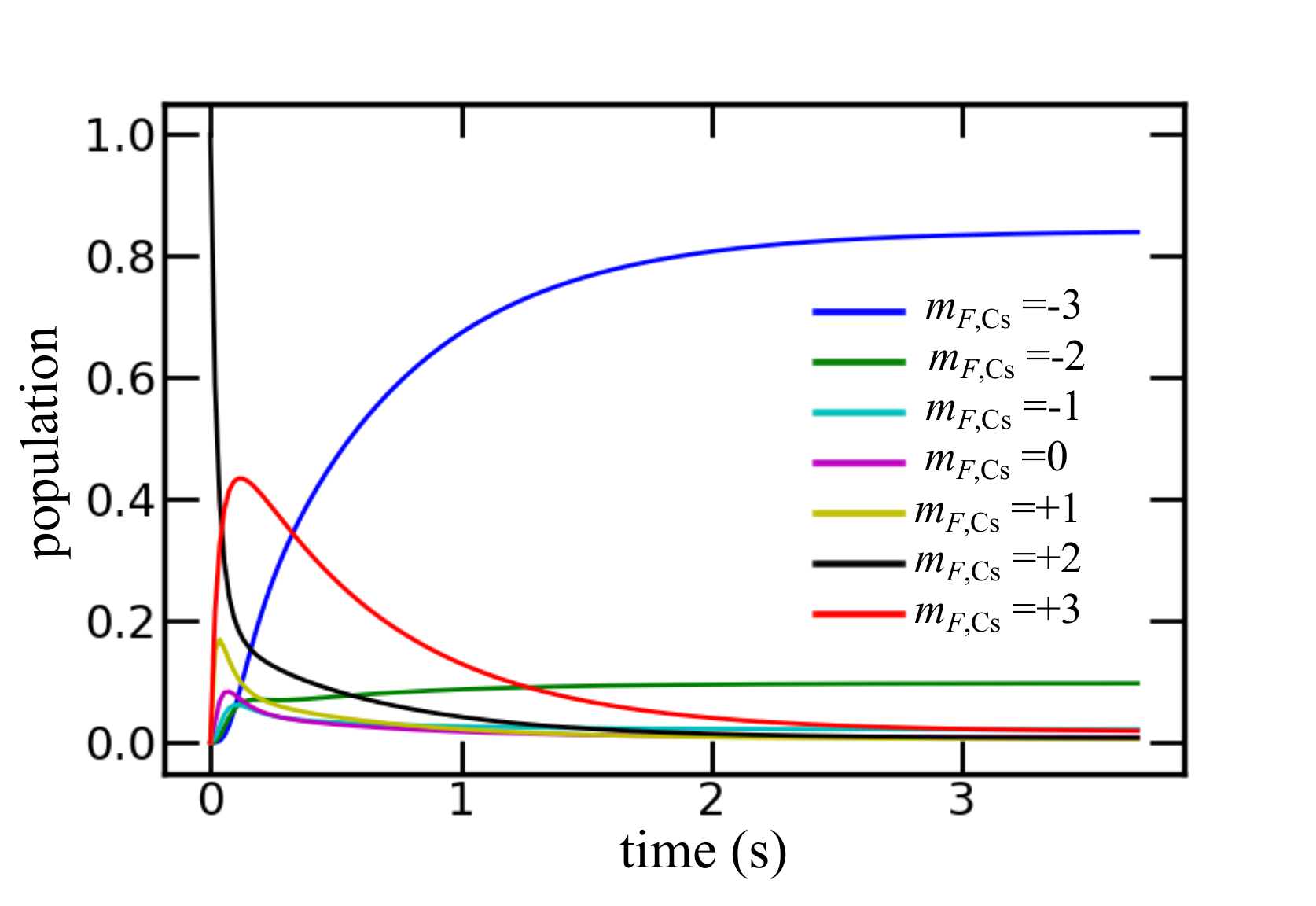}
	\end{center}
	\caption{Simulation of the spin states dynamic of Cs immersed into a Rb bath with $N_{\mathrm{Rb}}=7 \times 10^{3}$ atoms at $T$ = 400 nK and $B$ = 10 mG (expected endoergic fraction $p(B,T)=0.84$). The seven spin states ($m_{F,\mathrm{Cs}} \in [-3,-2,-1,0,1,2,3]$) are plotted. The steady-state is reached after almost 3 s interaction time.}
	\label{figure_theory_time_evolution}
\end{figure}

The spin population of the Cs atoms immersed into the Rb cloud ($F_{\mathrm{Rb}}=1, m_{F,\mathrm{Rb}}=0$) is governed by the endoergic and exoergic SE process. We model the population $N_{mF,\mathrm{Cs}}$ of the 7 internal states of the Cs atoms ($m_{F,\mathrm{Cs}} \in [-3,-2,-1,0,1,2,3]$) by a rate model. Each $m_{F,\mathrm{Cs}}$ spin state decays, on the one hand to the $m_{F+1,\mathrm{Cs}}$ state at the rate $\Gamma^{m_{F} \rightarrow m_{F+1}}$ due to the endoergic SE process and, on the other hand to the $m_{F-1,\mathrm{Cs}}$ state at the rate $\Gamma^{m_{F} \rightarrow m_{F-1}}$ due to the exoergic SE process (see Fig \ref{figure_rate_supplementary}). In the meanwhile each $m_{F,\mathrm{Cs}}$ spin state gains population from $m_{F-1,\mathrm{Cs}}$ at the rate $\Gamma^{m_{F-1} \rightarrow m_{F}}$ due to endoergic SE process and from $m_{F+1,\mathrm{Cs}}$ at the rate $\Gamma^{m_{F+1} \rightarrow m_{F}}$ due to exoergic SE process (see also Fig \ref{figure_rate_supplementary}). It translates to the following differential equation for each $N_{mF,\mathrm{Cs}}$ spin state population

\begin{equation} \label{3}
\begin{split}
  \dot{N}_{m_{F,\mathrm{Cs}}}  = & + \Gamma^{m_{F+1} \rightarrow m_{F}} N_{m_{F+1},\mathrm{Cs}}  \\
 &+ \Gamma^{m_{F-1} \rightarrow m_{F}} N_{m_{F-1},\mathrm{Cs}} \\
 &-(\Gamma^{m_{F} \rightarrow m_{F-1}}+\Gamma^{m_{F} \rightarrow m_{F+1}})N_{m_{F},\mathrm{Cs}}
\end{split}
\end{equation}

\noindent In order to solve these differentials equations, the different collisions rates $\Gamma_{i}$ have to be inferred (6 for the endoergic process and 6 for the exoergic process). They are given by
\begin{equation} \label{6}
\Gamma_{i} = \braket{n} \sigma_{i}(B,T) \bar{v}
\end{equation}

\noindent where $\braket{n}$ is the Cs-Rb density overlap, $\sigma_{i}(B,T)$ the scattering cross section (which depedends on the considered state) and $\bar{v}$ the relative velocity between Rb and Cs atoms. To calculate these 3 parameters, we first assume full thermalization of the Cs atoms in the Rb bath at temperature $T$. The thermalization rate $\Gamma_{\mathrm{ther}}$ of a Cs atom is given by \cite{Mudrich2002} 
\begin{equation} \label{7}
\Gamma_{\mathrm{ther}} = \frac{\Gamma_{\mathrm{el}}}{3} \xi \frac{N_{\mathrm{Rb}} + N_{\mathrm{Cs}}}{N_{\mathrm{Rb}}}
\end{equation}

\noindent where $\Gamma_{\mathrm{el}}$ is the  scattering elastic collision rate and $\xi= \frac{4 m_{\mathrm{Rb}} m_{\mathrm{Cs}}}{(m_{\mathrm{Rb}} + m_{\mathrm{Cs}})^{2}}$ the reduction factor for momentum exchange in a Cs-Rb collision due to the mass imbalance. The thermalization of a single Cs atoms in a large Rb bath ($N_{\mathrm{Cs}} \ll N_{\mathrm{Rb}}$) yields a thermalization rate of $\Gamma_{\mathrm{ther}} \approx \Gamma_{\mathrm{el}}/3 $. Since the elastic rate is 10 times higher than the SE rates, the thermalisation of the Cs atom is always ensured at the moment of the SE collisions. As a consequence, the relative velocity between Rb and Cs writes
\begin{equation} \label{8}
\bar{v} = \sqrt{\frac{8 k_{B}T}{\pi \mu}}
\end{equation}

\noindent where $\mu$ is the reduced mass. The density-density overlap $\braket{n}$ of Cs and Rb at density $n_{Cs}$ and $n_{Rb}$ is
\begin{equation} \label{9}
 \braket{n} = \int n_{Cs}(\vec{r}) n_{Rb}(\vec{r}) d\vec{r}
\end{equation}

\noindent and is calculated assuming a Maxwell-Boltzmann distribution for Cs and Rb.  Finally the different scattering cross sections are averaged over a thermalized distribution $\sigma_{i}(B,T) = \int p(E_{c}) \sigma_{i}(B,E_{c}) dE_{c}$, as explained in the previous section. \\

Starting with an initial Cs population in the $m_{F,\mathrm{Cs}}=2$ state, we numerically integrate equation \eqref{3} and find excellent agreement between the theory and the experimental data, as illustrated in Fig.\ref{figure_time_evolution_data} (a) in the main text. Moreover, we also simulate with our model the Cs spin population with $N_{\mathrm{Rb}}=7 \times 10^{3}$ atoms at $T$ = 400 nK and $B$ = 10 mG, which are typical numbers in our experiment. The result is plotted in Fig. \ref{figure_theory_time_evolution}. We observe that the steady-state is reached after an interaction time of almost 3 s, which would lead to a non negligible loss of Cs atoms due to three-body recombination (Rb-Rb-Cs). This loss rate writes $\Gamma_{\mathrm{3body}}= L_{3} \braket{n^{2}}$, with $\braket{n^{2}}=\int n_{\mathrm{Rb}}^{2}(\vec{r}) n_{\mathrm{Cs}}(\vec{r}) d\vec{r}$ and $L_{3}=28(1) \times 10^{-26}$ $\mathrm{Hz}$ $\mathrm{cm^{-6}}$ \cite{Mayer2018}. The expected value of the rate of three-body losses is $\Gamma_{\mathrm{3body}}= 0.66$ Hz, leading to an expected lifetime of Cs $\tau = 1/\Gamma_{\mathrm{3body}} = 1.5$ s. Therefore a large fraction of Cs atoms should be lost when the steady-state is reached in our system.

\subsection{Mean and fluctuation of energy, probe entropy and number of spin collisions}

\begin{figure}[t]
	\begin{center}	
		\includegraphics[scale=0.55]{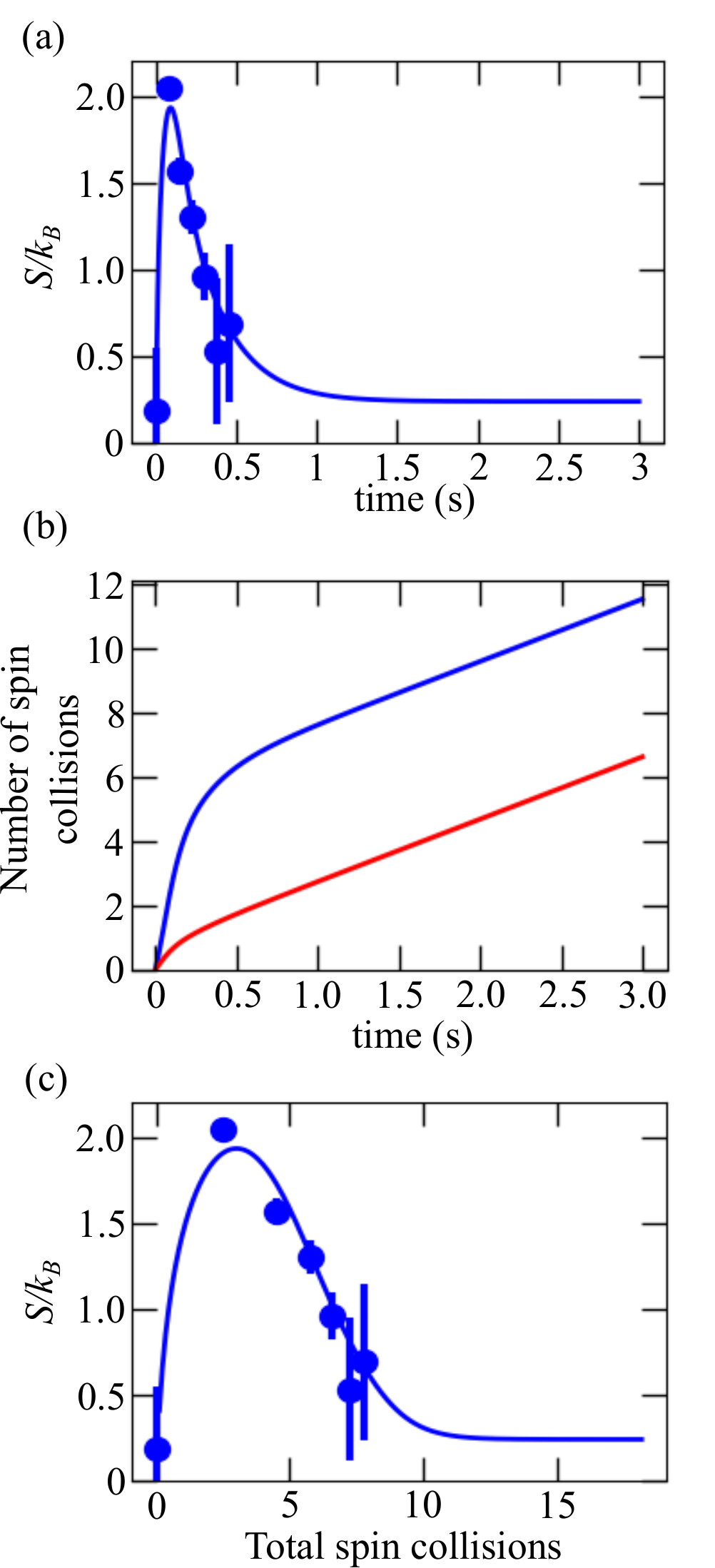}			
	\end{center}
	\caption{Evolution of the entropy $S$ of the probe. The Cs atoms are initially prepared in $m_{F,\mathrm{Cs}}=2$, immersed in a Rb bath at $T=$ 366 nK with $N_{\mathrm{Rb}}=6.7 \times 10^{3} $ atoms. The magnetic field is $B$ = 25 mG. Experimental points (circle) and theoretical predictions (solid lines) are shown. (a) Time evolution of the entropy of the probe $S$. (b) Time evolution of the mean number of spin collisions: in red the mean number of endoergic spin collisions and in blue the mean number of exoergic spin collisions. (c) The entropy of the probe in function of the total numbers of spin collisions is shown.
	}
	\label{figure_entropy_N_collisions_supplementary}
\end{figure}

We denote each of the seven Cs spin state populations with quantum number $m_{F} \in [3,2,1,0,-1,-2,-3]$ as $p(m_{F})$ with $\sum_{m_{F}} p_{m_{F}}=1$. From modelled quantum-state distributions, we can extract useful observables such as the mean energy $\braket{E}$, the variance of energy $\braket{E^{2}}$, the entropy $S$ of the Cs atom's spin population and the number of spin collisions. $\braket{E}$ is defined as 
\begin{equation} \label{10}
\braket{E} = \sum_{m_{F}=+3}^{m_{F}=-3} p(m_{F}) E_{m_{F}}
\end{equation}

\noindent where $E_{m_{F}} = (3-m_{F})\Delta E/2$, such as the energy of the ground state $\ket{m_{F}=3}$ is set to zero in our model. In the same way, $\braket{E^{2}}$ writes as
\begin{equation} \label{11}
\braket{E^{2}} = \sum_{m_{F}=+3}^{m_{F}=-3} p(m_{F}) E^{2}_{m_{F}}
\end{equation}

\noindent The fluctuations of energy is then given by $\sigma^{2}_{E} = \braket{E^{2}} - \braket{E}^{2}$. Finally the entropy $S$ is expressed as
\begin{equation} \label{12}
S= -k_{B} \sum_{m_{F}=+3}^{m_{F}=-3} p(m_{F}) \mathrm{log} (p(m_{F}))
\end{equation}

\noindent An example of time evolution of entropy $S$ is depicted in Fig.\ref{figure_entropy_N_collisions_supplementary} (a). We find that a maximum entropy is reached for only t= 90 ms, indicating that the nonequilibrium spin distribution can yield much better information. In order to quantify the the number of spin collisions necessary to reach the optimum, we first calculate the mean spin collisions rate $\braket{\Gamma(t)}$
\begin{equation} \label{13}
\begin{split}
  \braket{\Gamma(t)}   & = \braket{\Gamma_{\mathrm{endo}}(t)}+ \braket{\Gamma_{\mathrm{exo}}(t)}  \\
& = \sum_{m_{F}=+2}^{m_{F}=-3} p(m_{F}) \Gamma_{\mathrm{endo}}^{m_{F} \rightarrow m_{F}+1} \\
 &+ \sum_{m_{F}=+3}^{m_{F}=-2} p(m_{F}) \Gamma_{\mathrm{exo}}^{m_{F} \rightarrow m_{F}-1}
\end{split}
\end{equation}

\noindent The number of spin collisions $N_{\mathrm{spin}} = N_{\mathrm{endo}} + N_{\mathrm{exo}}$ is then deduced by integrating equation \eqref{13}
\begin{equation} \label{14}
\begin{split}
  N_{\mathrm{spin}}(t)   & =  N_{\mathrm{endo}} + N_{\mathrm{exo}}  \\
& = \int_{0}^{t}   (\braket{\Gamma_{\mathrm{endo}}(t')} + \braket{\Gamma_{\mathrm{endo}}(t')}) dt'  
\end{split}
\end{equation}

\noindent An example of time evolution of endoergic and exoergic spin collisions is represented in Fig. \ref{figure_entropy_N_collisions_supplementary} (b). We can then track the entropy of the probe in function of the number of spin collisions (Fig. \ref{figure_entropy_N_collisions_supplementary} (c)): we find that the maximum of entropy is obtained for only 3 mean spin collisions (2.5 mean exoergic collisions and 0.5 mean  endoergic spin collision).

\subsection{Steady-state}

\begin{figure}[t]
	\begin{center}	
		\includegraphics[scale=0.5]{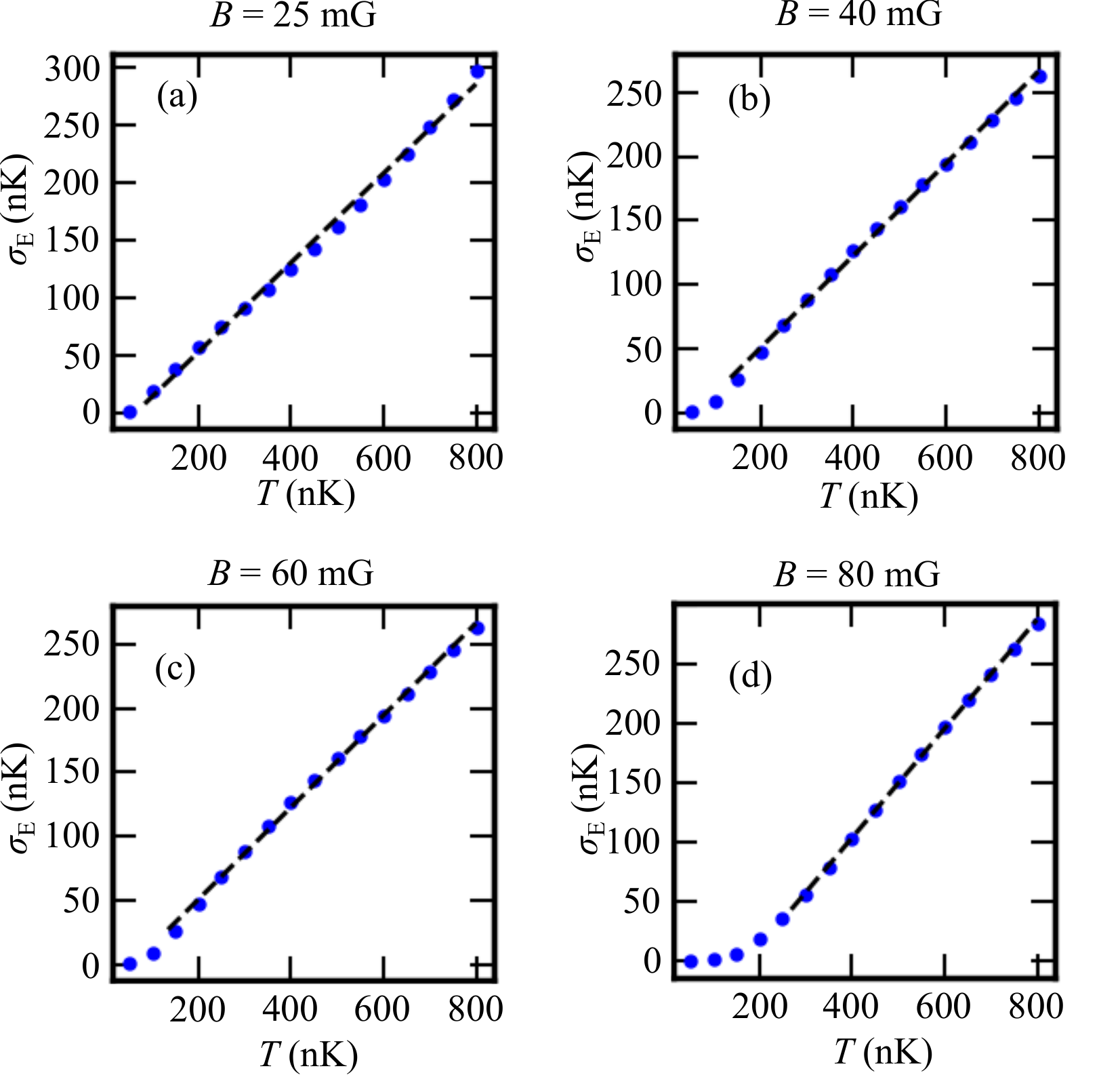}			
	\end{center}
	\caption{Fluctuation of energy $\sigma_{E}$ of the steady-state of the spin population of the probe in function of the temperature $T$ for a fixed magnetic field $B$: (a) $B$ = 25 mG, (b) $B$ = 40 mG, (c) $B$ = 60 mG, (d) $B$ = 80 mG. The dots are the theoretical points and the dashed lines represent the linear fit.
	}
	\label{figure_steady_state_supplementary}
\end{figure}

In the steady-state, the temperature only depends on the scattering cross section $\sigma_{i}(B,T)$ since all the rates $\Gamma_{i}$ have the same dependency in regards to the density $\braket{n}$ and the relative velocity $\bar{v}$ (equation \eqref{6}). Therefore thermometry can also be performed using the steady-state. To demonstrate this, we investigate the fluctuation of energy $\sigma_{E}$  to the steady-state for different temperature. The populations are inferred by solving equation \eqref{3} with $ \dot{N}_{m_{F},\mathrm{Cs}}=0$, replacing all the rates $\Gamma_{i}$ by the corresponding cross section $\sigma_{i}(B,T)$. Fig. \ref{figure_steady_state_supplementary} shows the behaviour of $\sigma_{E}$ with the temperature for different magnetic field $B$. If the thermal energy is significantly larger than the Zeeman energy and thus the fraction of endoergic SE amounts to more than a few percent according to equation \ref{2}, we observe a linear behaviour of the distribution's width with the temperature $T$. Furthermore we observe that the proportionality constant is independent to the magnetic field.  

\subsection{Extraction of the spin temperatures and spin magnetic fields}

\begin{figure}[t]
	\begin{center}	
		\includegraphics[scale=0.45]{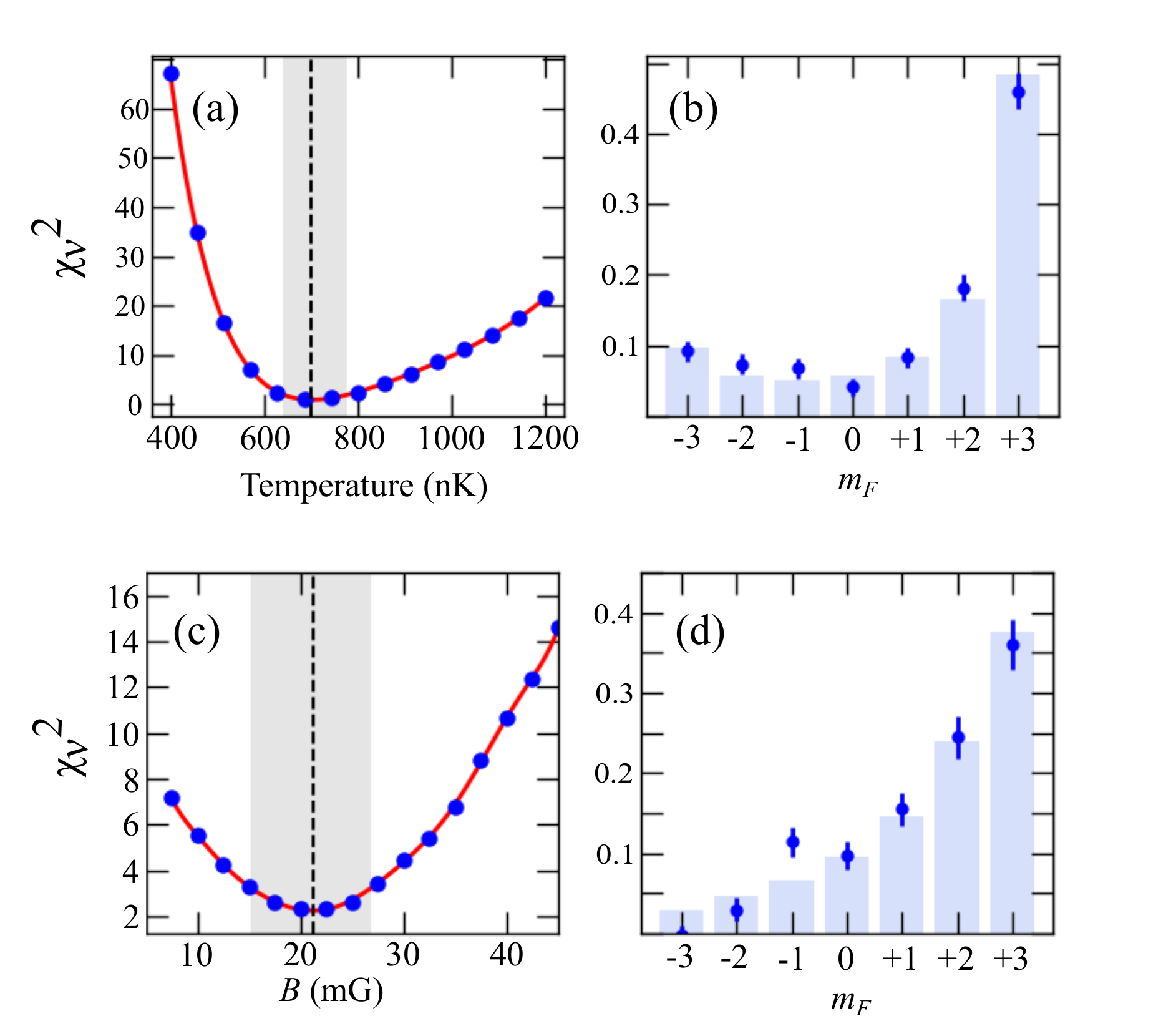}			
	\end{center}
	\caption{Extraction of the spin temperature $T_{\mathrm{spin}}$ and the spin magnetic field $B_{\mathrm{spin}}$ with a $\chi^{2}$-analysis. (a) shows a $\chi^{2}_{\nu} (T)$ curve (the magnetic field is fixed at $B$=10 mG) and (c) represents a $\chi^{2}_{\nu} (B)$ curve (the temperature is fixed at $T=1$ $\mu K$). In each curve, the vertical dashed line represents the minimum value of the $\chi^{2}_{\nu}$ which marks the extracted $T_{\mathrm{spin}}$ or $B_{\mathrm{spin}}$. The slightly grey band represents the 1 standard deviation, corresponding to an increase by 1 of $\chi^{2}_{\nu}$. We find $T_{\mathrm{spin}}=702^{+60}_{-76}$ for (a) and $B_{\mathrm{spin}}=21.6^{+5.7}_{-6.1}$ for (c). (b) and (d) represent the measured populations (dots) and the best parameter of interest with the model (histogram).}
	\label{figure_chi_squarre_supplementary}
\end{figure}

To extract the temperature or the magnetic field from the spin population of the probe, we perform a $\mathrm{\chi^{2}}$-analysis. For each measurement, comprising the seven internal states $m_{F,\mathrm{exp}} \in [-3,-2,1,0,1,2,3]$ of the Cs atom, a reduced $\chi^{2}_{\nu}$ is calculated

\begin{equation} \label{10}
\chi^{2}_{\nu} (\theta) = \frac{1}{\nu} \sum_{m_{F}} \frac{(P_{m_{F,\mathrm{exp}}}-P_{m_{F,\mathrm{theo}}}(\theta))^{2}}{\sigma_{m_{F,\mathrm{exp}}}^{2}}
\end{equation}

\noindent where $\theta= T$ or $B$, $P_{m_{F,exp}}$ the measured populations associated with the experimental errorbars $\sigma_{m_{F,\mathrm{exp}}}$. $P_{m_{F,\mathrm{theo}}}(\theta)$ are the theoretical populations deduced from our microscopic model, where only the parameter $\theta$ is a free parameter. Finally $\mathrm{\nu}$ is the degree of freedom, which is seven in our case. An example of $\mathrm{\chi^{2}}$-analysis is shown in Fig \ref{figure_chi_squarre_supplementary}. We extract the temperature $T_{\mathrm{spin}}$ (repectively the magnetic field $B_{\mathrm{spin}}$) by finding the minimum of $\chi^{2}_{\nu}(T)$ (repectively $\chi^{2}_{\nu}(B)$). The errorbar corresponds to the value of the parameter of interest $\theta = T$ or $B$ if we increase $\chi^{2}_{\nu}(\theta)$ by one, translating in 1 standard deviation $\mathrm{\sigma}$ in the errorbar \cite{Bevington2013}. In addition, we also study the systematic deviation of the spin temperatures $T_{\mathrm{spin}}$ due to the uncertainty of $\pm$ 2 mG on the magnetic field $B$. Including this effect in the $\mathrm{\chi^{2}}$-analysis, we find a systematic error close to 30 nK for the spin temperatures.

\subsection{Bures distance and Fisher information}

\begin{figure}[t]
	\begin{center}
		\includegraphics[scale=0.4]{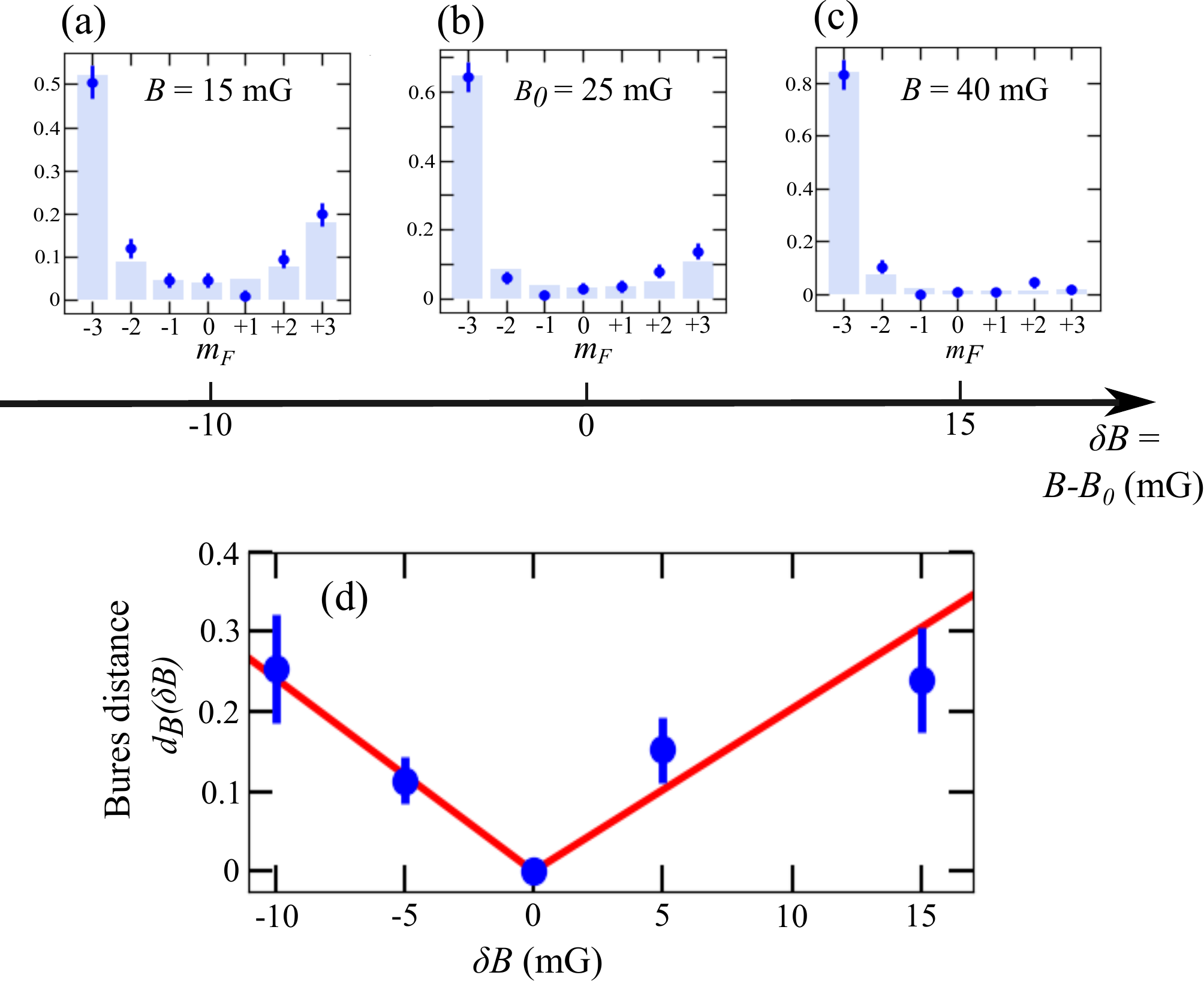}		
	\end{center}
	\caption{Experimental extraction of the Quantum Fisher information. The temperature is fixed at $T$ = 280 nK. In (a), (b) and (c) we represent the spin populations of the Cs atoms for different magnetic fields $B$: The dots are the experimental points and the histogram are the theoretical expectation values. In (d), we represent the Bures distance between the different states, centred at $B_{0}=25$ mG. In red, the Bures distance inferred by the theoretical spin populations and in blues by the experimental populations.}
	\label{figure_sensitivity_supplementary}
\end{figure}

The investigation on the thermal and magnetic sensitivity of our probe is done using the mathematical framework of the Quantum Fisher information. The thermal sensitivity means that the temperature $T$ is varied but the magnetic field $B$ is constant. On the contrary, the magnetic sensitivity means that $B$ is varied but $T$ is constant. Neglecting the coherences in the system, we describe each state by a diagonal density matrix $\hat{\rho}(B,T)$=$\sum_{m_{F}} P_{m_{F}}(B,T)  \ket{m_{F}} \bra{m_{F}}$, where $P_{m_{F}}(B,T)$ are the spin populations of the probe at $T$ and $B$. We denote the parameter of interest as $\theta$ (here $\theta=B$ or $T$). We quantify the distance between 2 quantum states at $\theta$ and $\theta + \delta \theta$ using Bures distance as \cite{Hiibner1992} \\

\begin{equation} \label{equation_Bures_distance_supplementary}
\begin{split}
d^{2}_{\mathrm{Bures}}(\delta \theta)   & =  2-2 \mathrm{tr} (\sqrt{\sqrt{\hat{\rho}(\theta+\delta \theta )} \hat{\rho}(\theta ) \sqrt{\hat{\rho}(\theta+\delta \theta )}})  \\
 &=  2-2 \sum_{m_{F}} \left[ P_{m_{F}}(\theta)P_{m_{F}}(\theta+\delta \theta ) \right]^{1/2}\\
\end{split}
\end{equation} 

\noindent The latter expression uses the fact that the density matrix is diagonal and thus the density operators between original and modified quantum state commute. In these conditions, the Bures distance coincides with the so-called Hellinger distance. The relation between the Bures distance and the Quantum Fisher information $F_{\theta}$ is \cite{Braunstein1994}

\begin{equation}
\label{Bures_distance_definition_B_supplementary}
d_{\mathrm{Bures}}(\delta \theta) = \sqrt{F_{\theta}}  \delta \theta \ + \ \mathcal{O} (\delta \theta^{2})
\end{equation}

\noindent In figure \ref{figure_sensitivity_supplementary} (d), we represent the Bures distance when $\theta=B$. We observe a linear behaviour of the Bures distance, the slopes thus representing the Fisher information, that we refer to the sensitivity. More precisely, we perform two Taylor expansions: one for $\delta \theta <0$ and one for $\delta \theta >0$. In general these two Taylor expansions are equal since the system considered is symmetric and  $d^{2}_{\mathrm{Bures}}(\delta \theta)$ is directly analysed \cite{Strobel12017}. However, in our case, due to the broken symmetry between endoergic and exoteric processes, these quantities are slightly different (between $10-20 \%$). Nonetheless they share the same behaviour. Hence there are no additional information in studying them separately and we simply study the mean value. 
		
\end{document}